\documentclass{optica-article}

\journal{opticajournal}
\articletype{Research Article}

\usepackage{amsmath,amssymb,bm}
\usepackage{graphicx}
\usepackage{booktabs}
\usepackage{siunitx}
\usepackage{tikz}
\usepackage{placeins}
\usetikzlibrary{arrows.meta,positioning,fit}

\begin{document}
    	
        		\title{Reconstructing intracell polarization with calibrated sapphire-window optics for
    			collimated blue-light generation in a rubidium cell}

\author{Sarah Ashby,\authormark{1} David Bostanjyan,\authormark{1} and Mikayel Khanbekyan\authormark{1,*}}
\address{\authormark{1}Institute for Physical Research, Ashtarak, Armenia}
\email{\authormark{*}khanbekyan@gmail.com}

\begin{abstract*}
We study the polarization dependence of collimated blue light (CBL) generated near 420~nm by resonant four-wave mixing in thermal rubidium vapor in a sapphire-window cell driven at 780 and 776~nm. Sapphire windows can transform the polarization of pump light before it reaches the vapor, complicating the interpretation of polarization-dependent nonlinear emission. We measured the 780-nm pump Stokes state at an external before-cell reference plane to calibrate the common HWP/QWP preparation optics and recorded 266 off-resonant external output Stokes states after the complete sapphire-window rubidium cell at 90, 100, and 110~$^\circ$C in two propagation orientations. 
Direct paired-wavelength measurements show a mean before-cell 780/776 separation of $4.29^\circ$ and temperature- and orientation-specific complete-cell rotations of $22.20^\circ$--$32.02^\circ$, with rms residuals of $4.83^\circ$--$12.87^\circ$. 
Using this calibrated co-prepared pump description, we analyzed 2574 measurements of total spectrally filtered 420-nm CBL intensity. Across the six temperature--orientation data sets, a complete nine-coefficient second-order response yielded cross-validated $R^2$ values of 0.636 to 0.828.
These paired-wavelength measurements establish, for each temperature--orientation data set, a fixed relative Stokes-space map between the two co-prepared pump trajectories, with the residual of this mapping defining the experimental uncertainty of the correlated two-pump reduction used in the response model.
Thus, the resulting framework combines a reconstructed vapor-entrance polarization coordinate with direct two-wavelength calibration of the assembled cell and provides a practical basis for polarization-controlled nonlinear-optical experiments in sapphire-window vapor cells.
\end{abstract*}

\section{Introduction}

Resonant atomic vapors provide an efficient medium for nonlinear frequency conversion because atomic resonance, optical pumping, and long-lived coherence can enhance light--matter interactions. In rubidium, two near-infrared fields near 780 and 776~nm drive the $5S_{1/2}\rightarrow5P_{3/2}\rightarrow5D_{5/2}$ ladder and generate a narrow directional field near 420~nm through resonant four-wave mixing~\cite{Zibrov2002,Meijer2006,Akulshin2009,Vernier2010}. The collimated blue-light (CBL) efficiency depends strongly on pump polarization because the driving fields select Zeeman excitation pathways and redistribute population among magnetic sublevels. Previous experiments have shown large differences between co-circular, counter-circular, and linear polarization configurations~\cite{Akulshin2009,Vernier2010}. More recently, polarization-structured excitation has been used to map polarization-resolved four-wave-mixing spectra in $^{85}$Rb~\cite{Wang2023}. The same wave-mixing cycle also involves the $5D_{5/2}\rightarrow6P_{3/2}$ transition near 5.23~$\mu$m, linking the blue output to the mid-infrared branch~\cite{Akulshin2017MidIR,Sebbag2019}.

Sapphire windows provide mechanical robustness and high-temperature compatibility for alkali-vapor cells~\cite{SarkisyanPapoyan1996,Sekiguchi2018}. They are also attractive when optical access to the mid-infrared branch is desired because refractive-index data for sapphire extend beyond 5.2~$\mu$m~\cite{Malitson1962}, while its thermal, mechanical, and chemical properties make it suitable for demanding vapor-cell geometries~\cite{Dobrovinskaya2009}. Sapphire is, however, a uniaxial crystal, so the windows of an assembled cell form part of the polarization optics. Crystal-axis orientation, bonding stress, and thermally induced stress can modify the polarization delivered to the vapor. Stress birefringence is therefore a known practical issue in sapphire vapor-cell construction~\cite{Sarkisyan2005}. Consequently, the polarization at the vapor entrance cannot in general be inferred from the externally prepared state and nominal window specifications alone.

This raises two related questions: how accurately can the external-to-external polarization transformation of the hot cell be calibrated, and how well can the resulting reconstructed polarization coordinate organize the measured CBL response? We address these questions in two experimentally distinct steps. Full-Stokes polarimetry is first used to characterize the optical transformation of the assembled cell. At 780~nm, the state prepared immediately before the oven, $\mathbf S_{\rm prep}$, and the state measured after the complete cell, $\mathbf S_{\rm out}^{\rm meas}$, are determined directly. A global lossless Jones-operator model is fitted to 266 measured off-resonant external output states at three temperatures and in both propagation orientations. Its effective entrance/exit factorization provides a reproducible reconstructed coordinate, $\mathbf S_{\rm atom}$, for the polarization at the vapor entrance. The distinction between externally measured and internally reconstructed states is maintained throughout the analysis. Stokes--Mueller methods provide a general framework for characterizing polarization-sensitive optical systems and nonlinear signals~\cite{Brasselet2011,Mazumder2012,Samim2015}.

The wavelength dependence of the assembled cell is then measured directly using matched 780- and 776-nm full-Stokes measurements at the same six preparation settings. The two wavelengths differ by $4.29^\circ$ on average at the before-cell reference plane, whereas the matched complete-cell output trajectories are related by best-fit temperature- and orientation-specific Poincar\'e-sphere rotations of approximately $22^\circ$--$32^\circ$. The residuals of these fits quantify the accuracy with which a single rigid rotation represents the paired-wavelength trajectory for each temperature--orientation data set. This paired calibration therefore characterizes the chromatic transformation of the complete assembled cell without assigning the wavelength dependence to either individual sapphire window.

With the optical reconstruction fixed, we then compare empirical response functions for the total spectrally filtered 420-nm CBL intensity, ranging from circularity and complete first-order Stokes terms to reduced and complete second-order bases. Repeated cross-validation and the Bayesian information criterion quantify their predictive performance~\cite{Schwarz1978}. For the co-prepared pump configuration studied here, the measured 780- and 776-nm trajectories form correlated polarization coordinates linked by the experimentally determined temperature- and orientation-specific rotations; the corresponding two-pump response reduction is developed in Sec.~\ref{sec-4}. Independent polarization control of the two pumps and polarization analysis of the generated CBL provide natural extensions of the present measurements.

The manuscript is organized as follows. Section~\ref{sec-2} describes the CBL experiment and the wavelength-resolved Stokes calibration. Section~\ref{sec-3} introduces the effective Jones reconstruction and the vapor-entrance coordinate. Section~\ref{sec-4} defines the empirical CBL-response models, the correlated two-pump reduction, and the model-selection procedure. Section~\ref{sec-5} presents the measured maps and model comparison, Sec.~\ref{sec-6} discusses the interpretation and limitations, and Sec.~\ref{sec-7} gives the conclusions and outlook.

\section{Experimental arrangement}
\label{sec-2}

\begin{figure}[!htbp]
    \centering
    \includegraphics[width=0.95\linewidth]{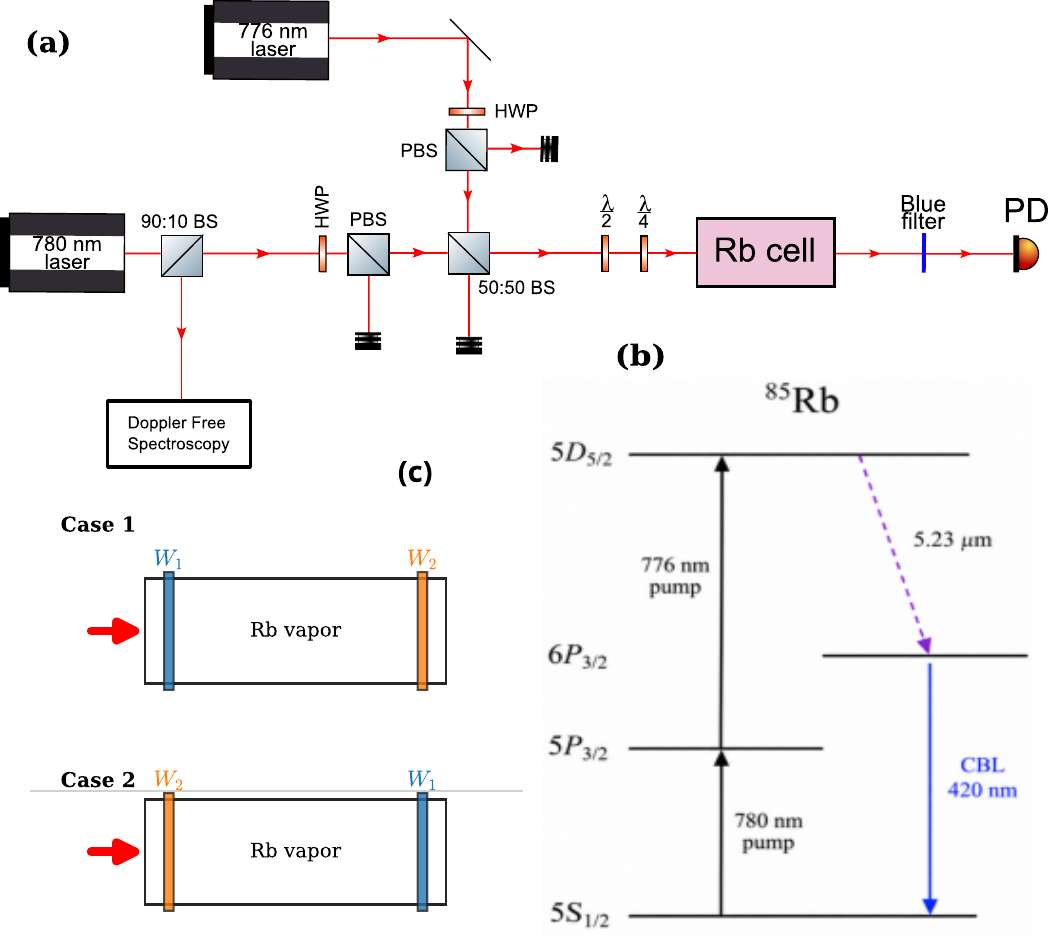}
    \caption{Experimental arrangement, atomic scheme, and cell orientations.
    (a) The 780- and 776-nm beams pass through separate HWP--PBS power-control stages and are combined at a 50:50 beam splitter. The co-propagating pumps then traverse the common HWP/QWP polarization-preparation optics before the heated Rb cell. A 90:10 beam splitter directs part of the 780-nm light to Doppler-free spectroscopy. The generated blue light is selected by a blue filter and detected with a photodetector.
    (b) Four-level scheme of the $^{85}$Rb wave-mixing process. The 780- and 776-nm pumps drive the $5S_{1/2}\rightarrow5P_{3/2}\rightarrow5D_{5/2}$ ladder; the cycle includes the $5D_{5/2}\rightarrow6P_{3/2}$ transition near 5.23~$\mu$m and produces the 420-nm CBL field on the $6P_{3/2}\rightarrow5S_{1/2}$ transition.
    (c) The two cell orientations exchange the entrance and exit sapphire windows, giving $W_1\rightarrow W_2$ and $W_2\rightarrow W_1$.}
    \label{fig:setup}
\end{figure}

\subsection{Collimated blue-light generation}

We employ an in-house-fabricated, sealed-off, 3-cm-long T-shaped glass cell with sapphire windows and a side arm containing natural rubidium. The vapor volume contains natural Rb, no buffer gas and no wall coating. The two circular technical-grade sapphire ($\mathrm{Al_2O_3}$) windows are bonded to the glass body, with each window of 17~mm in diameter and 1~mm thickness.  We assume that no crystallographic cut or optic-axis specification was provided for the technical-grade sapphire windows. Moreover, the nominal room-temperature specification of the individual windows would not by itself determine their polarization transformation in the assembled cell, because bonding to the glass body and operation at high temperatures can modify the effective birefringence, optical axes, and retardance. 

Two continuous-wave laser fields near 780 and 776~nm drive the $5S_{1/2}\rightarrow5P_{3/2}\rightarrow5D_{5/2}$ excitation ladder. The subsequent $5D_{5/2}\rightarrow6P_{3/2}$ and $6P_{3/2}\rightarrow5S_{1/2}$ transitions correspond to radiation near 5.23~$\mu$m and 420~nm, respectively, completing the four-wave-mixing cycle shown in Fig.~\ref{fig:setup}(b). The generated fields obey the corresponding energy-conservation relation, while phase matching selects the directional blue output.

The two pump beams emerge from their HWP--PBS power-control stages with the same linear polarization and are combined at the 50:50 beam splitter shown in Fig.~\ref{fig:setup}(a). After combination, both wavelengths traverse the same 1-inch mounted polymer zero-order half-wave plate (Thorlabs WPH10ME-780) and quarter-wave plate (Thorlabs WPQ10ME-780), both specified for 780~nm. The common HWP/QWP angles therefore vary the two pump states together. Any wavelength-dependent transformation introduced by the upstream combining optics is included in the directly measured before-cell Stokes calibration. The beams are collinear and co-propagating through the cell.

Beam-profiler measurements at the cell give $D4\sigma$ diameters of approximately 2.0 and 2.4~mm for the 780-nm beam and 1.9 and 2.2~mm for the 776-nm beam along the two principal transverse axes. Typical powers measured immediately before the cell, after the polarization-preparation optics, are approximately 20~mW at 780~nm and 2.7~mW at 776~nm. For an elliptical Gaussian with $D4\sigma=2w$ along each principal axis, these values correspond to Gaussian-equivalent peak intensities of approximately 1.06~W~cm$^{-2}$ at 780~nm and 0.165~W~cm$^{-2}$ at 776~nm. The 780-nm value is approximately $6.4\times10^2$ times the standard 1.67~mW~cm$^{-2}$ saturation intensity of the closed $^{85}$Rb D$_2$ cycling transition~\cite{Steck85}; in the thermal multilevel vapor this ratio serves as a reference for the strong-driving regime. For the 776-nm upper leg, the relevant saturation scale is set by the driven $5P_{3/2}$ population, Zeeman coupling strengths, and the $5P$--$5D$ coherence decay, so the measured peak intensity provides the direct experimental scale used here.

The temperature of the Rb-containing side arm was stabilized to approximately $\pm0.5~^\circ$C at 90, 100, or 110~$^\circ$C. These temperatures correspond to total Rb number densities of approximately $3.1\times10^{12}$, $6.0\times10^{12}$, and $1.12\times10^{13}$~cm$^{-3}$, respectively~\cite{Alcock1984}. The cell body was maintained approximately 10~$^\circ$C hotter than the side arm to keep the optical surfaces clear. Throughout the paper, the quoted temperature is the side-arm temperature. No external magnetic field or zero-field compensation was applied.

For the resonant CBL measurements, the 780-nm laser is locked to the $^{85}$Rb $5S_{1/2}(F=3)\rightarrow5P_{3/2}(F'=4)$ cycling transition using Doppler-free spectroscopy as the frequency reference. Before each polarization map, the 776-nm laser is scanned to locate the frequency producing the maximum CBL signal; that frequency is then held fixed throughout the map. The generated 420-nm light is selected with a blue optical filter of 10-nm bandwidth and detected with an OPT101 photodiode-based in-house-built optical detector with integrated transimpedance amplifier (a nominal bandwidth 14~kHz and rise time approximately 28~$\mu$s), so that the recorded CBL observable is the total collected, spectrally filtered 420-nm detector signal.

For the off-resonant Stokes measurements, the 780-nm frequency is displaced by several hundred megahertz from the nearby $^{87}$Rb resonances, at a frequency between the $F=1$ and $F=2$ transition groups in the Doppler-free reference spectrum. This setting places the polarimetry in a regime dominated by the optical transformation of the assembled cell. The corresponding 776-nm wavelength-resolved calibration uses the same optical geometry and analyzer procedure. At a side-arm temperature of 100~$^\circ$C, the complete-cell power transmission ratios measured under this off-resonant polarimetry condition are approximately 0.61 (780~nm) and 0.68 (776~nm) for $W_1\rightarrow W_2$, and 0.58 (780~nm) and 0.68 (776~nm) for $W_2\rightarrow W_1$. These ratios refer exclusively to the off-resonant calibration condition and provide the propagation scale associated with the Stokes measurements; the resonant CBL measurements use their own pump powers and atomic propagation conditions.

The two sapphire-window sequence orders are obtained by rotating the complete cell between the two working orientations. Relative to one another, these orientations differ by $180^\circ$ and exchange the entrance and exit sapphire windows, giving $W_1\rightarrow W_2$ and $W_2\rightarrow W_1$, Fig.~\ref{fig:setup}(c).

\subsection{Polarization preparation and analysis}

The input polarization is prepared using a half-wave plate (HWP) followed by a quarter-wave plate (QWP).
For each preparation setting the polarization is characterized from six analyzer projections: horizontal ($H$), vertical ($V$), diagonal ($D$), antidiagonal ($A$), right-circular ($R$), and left-circular ($L$). The normalized Stokes components are
\begin{align}
S_1 &= \frac{I_H-I_V}{I_H+I_V},\\
S_2 &= \frac{I_D-I_A}{I_D+I_A},\\
S_3 &= \frac{I_R-I_L}{I_R+I_L}.
\end{align}
The six analyzer projections determine one complete Stokes vector. Direct before-cell projections establish the external preparation at the reference plane in front of the oven, and direct after-cell projections establish the output state after the complete assembled cell. The vapor-entrance state used below is obtained by Jones-model reconstruction from these external measurements.

The dense HWP/QWP grid serves a different purpose from the six analyzer projections: it samples the nonlinear CBL response throughout the experimentally accessible polarization space. For $W_1\rightarrow W_2$, 13 HWP settings and 34 QWP settings give 442 CBL points per temperature. For $W_2\rightarrow W_1$, 13 HWP settings and 32 QWP settings give 416 points per temperature. In total, 2574 pump-polarization settings are analyzed. The global 780-nm cell reconstruction uses 266 complete off-resonant external output Stokes states spanning the three temperatures and two propagation orientations. Separate direct before-cell Stokes measurements calibrate the HWP/QWP preparation optics.

\subsection{Wavelength-resolved 780/776-nm polarimetry}
\label{sec:wavelength-data}

\begin{table}[!htbp]
	\centering
	\caption{Measured wavelength dependence of the complete assembled cell. The rotation angle characterizes the best proper Poincar\'e rotation mapping the matched 780-nm external output states onto the corresponding 776-nm states for each temperature and orientation. The rms residual is the angular mismatch after that rotation.}
	\label{tab:wavelength}
	\small
	\begin{tabular}{crr}
		\toprule
		$T$ ($^\circ$C), orientation & rotation angle & rms residual \\
		\midrule
		90, $W_1\!\rightarrow\!W_2$  & $29.39^\circ$ & $4.83^\circ$ \\
		90, $W_2\!\rightarrow\!W_1$  & $23.79^\circ$ & $5.57^\circ$ \\
		100, $W_1\!\rightarrow\!W_2$ & $29.09^\circ$ & $5.44^\circ$ \\
		100, $W_2\!\rightarrow\!W_1$ & $23.15^\circ$ & $7.80^\circ$ \\
		110, $W_1\!\rightarrow\!W_2$ & $32.02^\circ$ & $7.20^\circ$ \\
		110, $W_2\!\rightarrow\!W_1$ & $22.20^\circ$ & $12.87^\circ$ \\
		\bottomrule
	\end{tabular}
\end{table}

The pump-wavelength dependence is measured directly with matched Stokes measurements at 780 and 776~nm. Six well-separated preparation settings spanning linear and circular sectors are measured before the cell and after the complete cell. The after-cell measurements are repeated at 90, 100, and 110~$^\circ$C for both $W_1\rightarrow W_2$ and $W_2\rightarrow W_1$. The same $H,V,D,A,R,L$ analyzer procedure is used at the two wavelengths, so the comparison is made directly in normalized Stokes space.

For two normalized Stokes vectors $\mathbf S_a$ and $\mathbf S_b$, we use the Poincar\'e-sphere angular separation
\begin{equation}
\Gamma(\mathbf S_a,\mathbf S_b)=
\cos^{-1}\!\left(\mathbf S_a\cdot\mathbf S_b\right).
\label{eq:wavelength-angle}
\end{equation}
Across the six before-cell preparation states, the mean 780/776 angular separation is $4.29^\circ$ and the largest separation is $5.71^\circ$. Thus the common zero-order wave-plate pair prepares closely matched states at the two pump wavelengths at the external reference plane.

To characterize the additional wavelength dependence introduced by the assembled cell, for each temperature and orientation we determine the proper Poincar\'e rotation $R_{780\rightarrow776}^{(T,d)}\in SO(3)$ that gives the least-squares mapping between the six matched external output states,
\begin{equation}
\mathbf S_{\rm out}^{776}
\simeq
R_{780\rightarrow776}^{(T,d)}
\mathbf S_{\rm out}^{780}.
\label{eq:wavelength-rotation}
\end{equation}
Each temperature--orientation data set is calibrated independently with its own measured transformation. Table~\ref{tab:wavelength} summarizes the fitted rotation angle and the rms angular residual for the six measured conditions.

The wavelength-resolved measurements therefore separate two effects experimentally. 
At the before-cell reference plane, the mean 780/776 mismatch is only a few degrees, consistent with the weak wavelength dependence expected for the zero-order preparation wave plates over the 4-nm separation between the two pumps. 
After propagation through the assembled sapphire-window cell, the relative transformation is substantially larger. 
Within each propagation direction, the fitted rotation angle varies only modestly with temperature: from $29.09^\circ$ to $32.02^\circ$ for $W_1\rightarrow W_2$ and from $22.20^\circ$ to $23.79^\circ$ for $W_2\rightarrow W_1$. 
This weak temperature dependence supports the interpretation that the dominant effect originates from the sapphire-window assembly rather than from the vapor itself. 
Five of the six conditions have rotation-fit rms residuals between $4.83^\circ$ and $7.80^\circ$; the 110~$^\circ$C, $W_2\rightarrow W_1$ condition gives $12.87^\circ$. 
These residuals define the experimental accuracy of the rigid-rotation representation used below. 
The complete-cell transformation is the directly measured quantity in this wavelength comparison, while the individual-window decomposition is retained as an effective model coordinate in the 780-nm reconstruction.

\FloatBarrier

\section{Polarization reconstruction of the sapphire cell}
\label{sec-3}

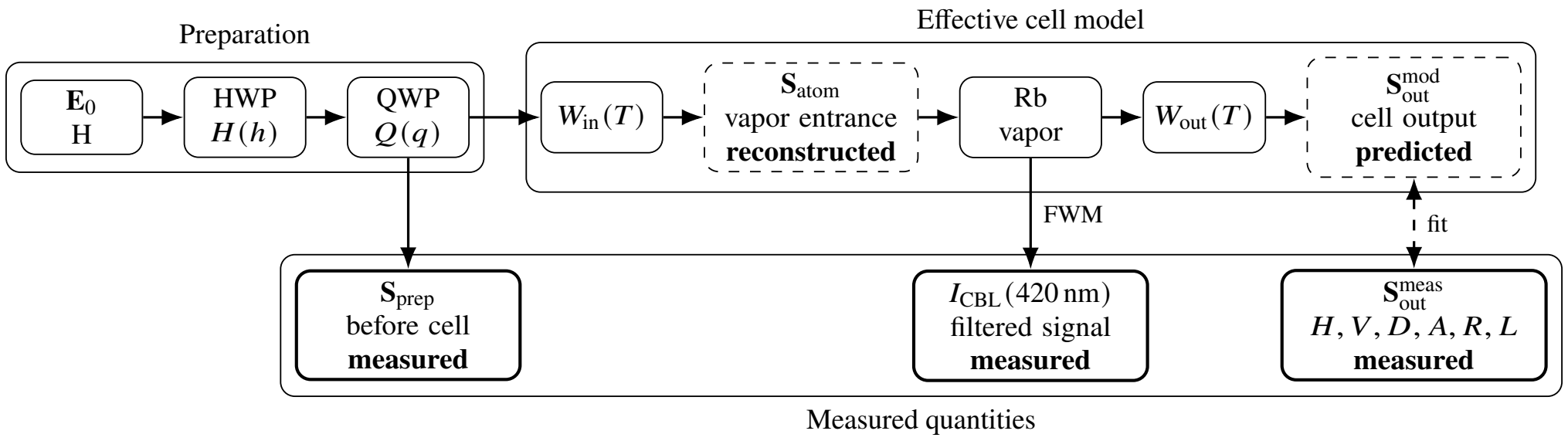
\begin{figure}[!htbp]
	\centering
	\resizebox{\linewidth}{!}{%
		\begin{tikzpicture}[
			font=\footnotesize,
			node distance=3mm and 4mm,
			>=Latex,
			flow/.style={-Latex, line width=0.7pt},
			comp/.style={<->, dashed, line width=0.7pt},
			box/.style={
				draw, rounded corners, align=center,
				inner sep=3pt, minimum height=7mm
			},
			meas/.style={box, line width=0.9pt},
			model/.style={box, dashed},
			op/.style={box, minimum width=12mm},
			group/.style={draw, rounded corners, inner sep=4pt}
			]
			
			%---------------- top row: model / optical path
			\node[box, minimum width=12mm] (e0) {$\mathbf E_0$\\H};
			\node[op, right=of e0] (hwp) {HWP\\$H(h)$};
			\node[op, right=of hwp] (qwp) {QWP\\$Q(q)$};
			
			\node[op, right=7mm of qwp] (winin) {$W_{\rm in}(T)$};
			\node[model, right=of winin, text width=19mm] (atom)
			{$\mathbf S_{\rm atom}$\\
				vapor entrance\\
				\textbf{reconstructed}};
			\node[box, right=of atom, minimum width=14mm] (vapor) {Rb\\vapor};
			\node[op, right=of vapor] (winout) {$W_{\rm out}(T)$};
			\node[model, right=of winout, text width=19mm] (outmod)
			{$\mathbf S_{\rm out}^{\rm mod}$\\
				cell output\\
				\textbf{predicted}};
			
			\draw[flow] (e0) -- (hwp);
			\draw[flow] (hwp) -- (qwp);
			\draw[flow] (qwp) -- (winin);
			\draw[flow] (winin) -- (atom);
			\draw[flow] (atom) -- (vapor);
			\draw[flow] (vapor) -- (winout);
			\draw[flow] (winout) -- (outmod);
			
			%---------------- bottom row: measurements
			\node[meas, below=11mm of qwp, text width=20mm] (prep)
			{$\mathbf S_{\rm prep}$\\
				before cell\\
				\textbf{measured}};
			
			\node[meas, below=9mm of outmod, text width=24mm] (outmeas)
			{$\mathbf S_{\rm out}^{\rm meas}$\\
				$H,V,D,A,R,L$\\
				\textbf{measured}};
			
			\node[meas, below=11mm of vapor, text width=21mm] (cbl)
			{$I_{\rm CBL}(420\,\mathrm{nm})$\\
				filtered signal\\
				\textbf{measured}};
			
			%---------------- vertical / comparison links
			\draw[flow] (qwp) -- (prep);
			\draw[comp] (outmod) -- node[right,font=\scriptsize]{fit} (outmeas);
			\draw[flow] (vapor) -- node[right,font=\scriptsize]{FWM} (cbl);
			
			%---------------- grouping boxes
			\node[group, fit=(e0)(hwp)(qwp),
			label={[font=\footnotesize]above:Preparation}] {};
			
			\node[group, fit=(winin)(atom)(vapor)(winout)(outmod),
			label={[font=\footnotesize]above:Effective cell model}] {};
			
			\node[group, fit=(prep)(outmeas)(cbl),
			label={[font=\footnotesize]below:Measured quantities}] {};
			
		\end{tikzpicture}%
	}
	\caption{Schematic of the polarization reconstruction and CBL-response workflow.
		Solid boxes denote directly measured quantities; dashed boxes denote reconstructed or model-predicted quantities.
		The before-cell state $\mathbf S_{\rm prep}$ is measured at the external reference plane after the HWP/QWP preparation optics, and the analyzer after the complete cell provides $\mathbf S_{\rm out}^{\rm meas}$. The global 780-nm Jones model predicts $\mathbf S_{\rm out}^{\rm mod}$ and supplies the model-reconstructed vapor-entrance coordinate $\mathbf S_{\rm atom}$ through its effective entrance-window factor. The independent resonant measurement records the total spectrally filtered 420-nm CBL intensity as a scalar detector signal.}
	\label{fig:workflow}
\end{figure}

The optical model determines how the polarization prepared by the external HWP/QWP pair is transformed by the assembled sapphire cell before the CBL data are analyzed. Here and thereafter, we use three canonical polarization states: $\mathbf S_{\rm prep}$ is the state immediately before the cell, $\mathbf S_{\rm atom}$ is the reconstructed state immediately after the entrance sapphire window, that is at the entrance to the atomic vapor, and $\mathbf S_{\rm out}^{\rm mod}$ is the model-predicted state after the exit window. The CBL response models in Sec.~4 use $\mathbf S_{\rm atom}$ as their polarization coordinate. Again, only the off-resonant Stokes measurements are used to determine the optical-model parameters (see Fig.~\ref{fig:workflow}, for details).

\begin{figure}[!htbp]
	\centering
	\includegraphics[width=0.8\linewidth]{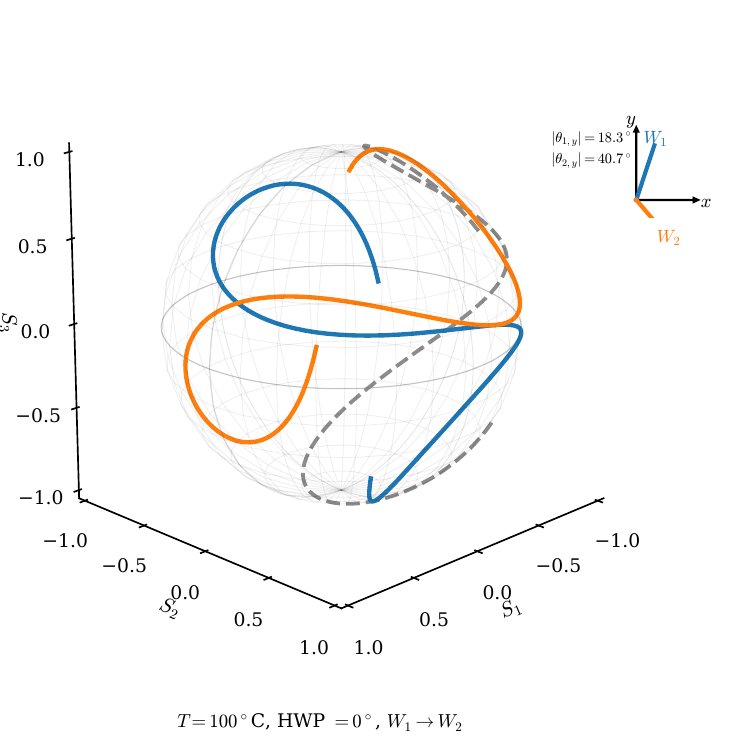}
	\caption{Poincar\'e-sphere representation of the model-reconstructed 780-nm polarization evolution for $T=100~^\circ$C, HWP$=0^\circ$, and the $W_1\rightarrow W_2$ orientation. The QWP angle varies continuously from $-80^\circ$ to $75^\circ$. The gray dashed curve is $\mathbf S_{\rm prep}$ before the cell, the blue solid curve is the model-reconstructed $\mathbf S_{\rm atom}$ immediately after the entrance window $W_1$, and the orange solid curve is the model-predicted $\mathbf S_{\rm out}^{\rm mod}$ after $W_2$. For this example, one representative factorization uses $(\theta,\alpha,\delta) = (-18.31^\circ,-43.72^\circ,-41.95^\circ)$ for $W_{1f}$ and $(-49.25^\circ,-118.08^\circ,168.46^\circ)$ for $W_{2f}$. Table~S3 gives the complete representative parameter set.}
	\label{fig:poincare}
\end{figure}

For the initially horizontal field the input Jones vector reads
\begin{equation}
\mathbf E_0=\begin{pmatrix}1\\0\end{pmatrix} .
\end{equation}
Then, the calibrated preparation optics generate
\begin{equation}
\mathbf E_{\rm prep}(h,q)=
\mathcal N\!\left[Q(q)H(h)\mathbf E_0\right],
\label{eq:prep}
\end{equation}
where $H$ and $Q$ include the calibrated HWP/QWP retardances and effective angular offsets, and $\mathcal N[\mathbf E]=\mathbf E/\lVert\mathbf E\rVert$. 

In general, a linear retarder with effective axis angle $\theta$ and retardance $\delta$ is represented by
\begin{equation}
J(\theta,\delta)=
R(\theta)
\begin{pmatrix}
e^{-i\delta/2} & 0\\
0 & e^{+i\delta/2}
\end{pmatrix}
R(-\theta),
\label{eq:retarder}
\end{equation}
where $R(\theta)$ is a two-dimensional Jones rotation. Each effective window operator can be written as
\begin{equation}
W_j(T)=R(\alpha_j)J\!\left[\theta_j,\delta_j(T)\right].
\label{eq:window}
\end{equation}
Here $\theta_j$ is an effective birefringent-axis angle, $\alpha_j$ is an additional effective rotation, and $\delta_j(T)$ is the temperature-dependent retardance. This three-parameter unitary operator represents the deterministic polarization transformation associated with an assembled sapphire window in a given cell orientation.

For the $W_1\rightarrow W_2$ orientation,
\begin{align}
\mathbf E_{\rm atom}^{(1)}&=\mathcal N\!\left[W_{1f}(T)\mathbf E_{\rm prep}\right],\\
\mathbf E_{\rm out}^{(1)}&=\mathcal N\!\left[W_{2f}(T)W_{1f}(T)\mathbf E_{\rm prep}\right],
\label{eq:case1}
\end{align}
and for the $W_2\rightarrow W_1$ orientation,
\begin{align}
\mathbf E_{\rm atom}^{(2)}&=\mathcal N\!\left[W_{2b}(T)\mathbf E_{\rm prep}\right],\\
\mathbf E_{\rm out}^{(2)}&=\mathcal N\!\left[W_{1b}(T)W_{2b}(T)\mathbf E_{\rm prep}\right].
\label{eq:case2}
\end{align}
The normalized Stokes vector reconstructed at the vapor entrance is denoted
$\mathbf S_{\rm atom}$, while the normalized Stokes vector predicted by the
model after the complete cell is denoted $\mathbf S_{\rm out}^{\rm mod}$.
The external analyzer provides $\mathbf S_{\rm out}^{\rm meas}$, which is compared directly with $\mathbf S_{\rm out}^{\rm mod}$ in the global 780-nm fit. Thus, $\mathbf S_{\rm prep}$ and $\mathbf S_{\rm out}^{\rm meas}$ are measured externally, whereas $\mathbf S_{\rm atom}$ is the vapor-entrance coordinate reconstructed from the fitted factorization.

The four effective operators $W_{1f}$, $W_{2f}$, $W_{2b}$, and $W_{1b}$ contain four axis/rotation pairs $(\theta_j,\alpha_j)$. These eight angular parameters are shared by all three temperatures because they describe fixed orientations of the effective optical elements. Four retardances $\delta_j(T)$ are fitted at each temperature, giving twelve temperature-dependent retardances across the three-temperature data set. The global factorized model therefore contains 20 cell parameters and is fitted simultaneously to 266 measured external output Stokes states. A normalized Stokes vector contributes two independent coordinates, so the global data set provides 532 independent normalized polarization coordinates for the 20-parameter fit. Each effective operator is unitary, and the same operator acts on every preparation state for each temperature and orientation.

The external polarimetry constrains the complete cell transformation
$W_{\rm out}(T)W_{\rm in}(T)$ directly, whereas its decomposition into
separate entrance- and exit-window operators is not unique. In particular,
for any unitary operator $U$,
\[
W_{\rm in}\rightarrow U W_{\rm in},
\qquad
W_{\rm out}\rightarrow W_{\rm out}U^{-1},
\]
leaves the externally observable product
$W_{\rm out}W_{\rm in}$ unchanged, while rotating the intermediate
polarization coordinate assigned to the vapor entrance. Equivalent
descriptions also arise from the periodic parametrization of retarder angles
and retardances and from the physically irrelevant global Jones phase.
Accordingly, the fitted parameters of the individual windows should be
interpreted as one effective factorization of the measured complete-cell
transformation, rather than as uniquely determined physical properties of
each window. Table~S3 of Supplement~1 lists one representative parameter set
and summarizes these equivalences.

The global fit minimizes the angular separation between measured and model-predicted external Stokes states,
\begin{equation}
\Delta\Omega_i=\cos^{-1}\!\left[
\frac{\mathbf S_i^{\,\rm meas}\!\cdot\mathbf S_i^{\,\rm mod}}
{\lVert\mathbf S_i^{\,\rm meas}\rVert
 \lVert\mathbf S_i^{\,\rm mod}\rVert}
\right],
\label{eq:stokesangle}
\end{equation}
with
\begin{equation}
\Delta\Omega_{\rm rms}=
\left(\frac{1}{N}\sum_{i=1}^{N}\Delta\Omega_i^2\right)^{1/2}.
\label{eq:stokesrms}
\end{equation}
The final global 780-nm fit gives $\Delta\Omega_{\rm rms}=5.51^\circ$.
The rms value represents the aggregate angular mismatch between measured and reconstructed polarization states over the complete data set---it can arise from analyzer and wave-plate calibration uncertainty, spatial averaging across the finite beam and sapphire windows, weak polarization-dependent transmission outside the lossless model, and other residual structure of the off-resonant cell/polarimeter response. In principle, the remaining error provides a practical measure of the accuracy with which the effective Jones model represents the assembled cell.

Figure~\ref{fig:poincare} gives a representative geometrical view of the fitted transformation. At $T=100~^\circ$C and fixed HWP angle, continuously varying the QWP traces a curve on the Poincar\'e sphere. The gray dashed curve is $\mathbf S_{\rm prep}$, the blue curve is $\mathbf S_{\rm atom}$ after the entrance window $W_1$, and the orange curve is $\mathbf S_{\rm out}^{\rm mod}$ after $W_2$. The sapphire windows therefore displace and reorient the experimentally accessible polarization trajectory on the Poincar\'e sphere. The inset shows the corresponding fitted effective axes in the laboratory frame.

\FloatBarrier

\section{Empirical CBL response models}
\label{sec-4}

For the CBL analysis, the polarization coordinate
$\mathbf S_{\rm atom}=(S_1,S_2,S_3)$ denotes the model-reconstructed
780-nm vapor-entrance state, with
\begin{equation}
	S_{1}^2+S_{2}^2+S_{3}^2=1.
	\label{eq:stokesnorm}
\end{equation}
Each HWP/QWP setting simultaneously controls both pump polarizations. We
treat each temperature and propagation orientation as a separate CBL data
set. The coefficients of each empirical response model are fitted
independently within that data set, whereas the Jones parameters established
from off-resonant polarimetry remain fixed throughout this stage.

For each temperature $T$ and propagation orientation $d$, we use the
reconstructed 780-nm vapor-entrance Stokes vector,
$\mathbf S_{780}\equiv\mathbf S_{\rm atom}$, as the working polarization
coordinate. For the co-prepared pumps, we represent the corresponding
776-nm vapor-entrance trajectory by a fixed rotation for each temperature and orientation,
\begin{equation}
	\mathbf S_{776}
	\approx
	\mathcal R_{T,d}\mathbf S_{780},
	\label{eq:two-pump-rotation}
\end{equation}
where $d$ labels the propagation orientation. The paired-wavelength
measurements support this fixed-rotation representation: at the directly
measured complete-cell output, the matched 780- and 776-nm trajectories
are described by single rigid rotations with rms residuals of
$4.83^\circ$--$12.87^\circ$. These measurements test the accuracy of the
rigid-trajectory approximation for each temperature and orientation, while the particular intracell
rotation $\mathcal R_{T,d}$ is not determined independently.

\subsection{Model $M_1$: Circularity and first-order}

Polarization-dependent CBL experiments show strong differences between circular and linear pump configurations because the pump polarization selects different Zeeman excitation pathways~\cite{Akulshin2009,Vernier2010,Wang2023}. We therefore use the magnitude of the circular Stokes component as the minimal benchmark,
\begin{equation}
M_1:\qquad I_{\mathrm{CBL}}=c_0+c_1|S_3|,
\label{eq:m1}
\end{equation}
which quantifies the part of the measured response associated with the magnitude of the circular Stokes component.

The most general first-order scalar response of the two co-prepared pump
polarizations is
\begin{equation}
	I_{\rm CBL}
	=
	c_0
	+\mathbf a^{\mathsf T}\mathbf S_{780}
	+\mathbf b^{\mathsf T}\mathbf S_{776}.
	\label{eq:two-pump-linear}
\end{equation}
For a fixed rotation, substitution of
Eq.~\eqref{eq:two-pump-rotation} gives
\begin{equation}
	I_{\rm CBL}
	=
	c_0+
	\left(
	\mathbf a+\mathcal R_{T,d}^{\mathsf T}\mathbf b
	\right)^{\mathsf T}
	\mathbf S_{780}.
	\label{eq:two-pump-linear-reduced}
\end{equation}
Thus a complete first-order response of the correlated two-pump
trajectory remains within the three-component linear Stokes basis:
\begin{equation}
	M_{\rm lin}:\qquad
	I_{\rm CBL}
	=
	c_0+c_1S_{1,780}+c_2S_{2,780}+c_3S_{3,780}.
	\label{eq:Mlin}
\end{equation}
This four-parameter model provides the complete linear response basis on the normalized Stokes sphere.
Note, the wavelength-dependent rotation changes the effective fitted
coefficients but does not increase the dimension of the complete
first-order basis.

Here and in the following, we retain the internal model
labels $M_1$, $M_{10}$, and $M_{13}$ used during development; the numerical
indices serve only as identifiers and have no particular physical meaning.

\subsection{Models $M_{10}$ and $M_{13}$: Reduced and complete second-order}

We start from the most general quadratic scalar function of the normalized Stokes vector:
\begin{align}
	I_{\rm CBL}
	={}&
	c_0
	+\mathbf a^{\mathsf T}\mathbf S_{780}
	+\mathbf b^{\mathsf T}\mathbf S_{776}
	\nonumber\\
	&+
	\mathbf S_{780}^{\mathsf T}
	A
	\mathbf S_{780}
	+
	\mathbf S_{776}^{\mathsf T}
	B
	\mathbf S_{776}
	+
	\mathbf S_{780}^{\mathsf T}
	C
	\mathbf S_{776},
	\label{eq:two-pump-quadratic}
\end{align}
where $A$, $B$, and $C$ are coefficient matrices. 
Substituting Eq.~\eqref{eq:two-pump-rotation},	
\begin{equation}
	I_{\rm CBL}
	=
	c_0
	+\mathbf c^{\mathsf T}\mathbf S_{780}
	+\mathbf S_{780}^{\mathsf T}
	Q_{\rm eff}
	\mathbf S_{780},
	\label{eq:quadratic-reduced}
\end{equation}
with
\begin{equation}
	\mathbf c
	=
	\mathbf a+\mathcal R_{T,d}^{\mathsf T}\mathbf b,
	\label{eq:ceff}
\end{equation}
and
\begin{equation}
	Q_{\rm eff}
	=
	A
	+\mathcal R_{T,d}^{\mathsf T}B\mathcal R_{T,d}
	+\frac{1}{2}
	\left(
	C\mathcal R_{T,d}
	+\mathcal R_{T,d}^{\mathsf T}C^{\mathsf T}
	\right).
	\label{eq:Qeff}
\end{equation}
Thus, the nonzero wavelength-dependent rotation is retained. Because of the normalization condition, Eq. \eqref{eq:stokesnorm}
one diagonal quadratic term is linearly dependent on the intercept, and the complete second-order response therefore contains nine independent
coefficients and can be written as
\begin{align}
	M_{13}:\qquad
	I_{\rm CBL}
	={}&
	c_0
	+c_1S_{1,780}
	+c_2S_{2,780}
	+c_3S_{3,780}
	\nonumber\\
	&+
	c_{11}S_{1,780}^2
	+c_{22}S_{2,780}^2
	+c_{12}S_{1,780}S_{2,780}
	\nonumber\\
	&+
	c_{13}S_{1,780}S_{3,780}
	+c_{23}S_{2,780}S_{3,780}.
	\label{eq:M13}
\end{align}
Accordingly, $M_{13}$ is the complete second-order scalar response of
the correlated two-pump trajectory expressed in the reconstructed
780-nm Stokes coordinate. In general, higher-order Stokes coordinates provide a general methodological framework for polarization-dependent nonlinear observables~\cite{Samim2015}.

For comparison, we also retain the reduced four-parameter second-order model
\begin{equation}
	M_{10}:\qquad
	I_{\rm CBL}
	=
	c_0
	+c_a\left(S_{1,780}^2+S_{2,780}^2\right)
	+c_bS_{2,780}S_{3,780}
	+c_cS_{1,780}S_{2,780}.
	\label{eq:M10}
\end{equation}
Because $M_{10}$ retains only selected quadratic combinations, it is
coordinate dependent: a rotation between the 780- and 776-nm Stokes
coordinates generally generates quadratic terms outside this restricted
basis.

\subsection{Fitting, cross-validation, and model selection}

For a given temperature and propagation direction, each model is written in the linear-regression form
\begin{equation}
	\mathbf y=X\boldsymbol{\beta}+\boldsymbol{\epsilon},
	\label{eq:linearfit}
\end{equation}
where $\mathbf y$ contains the measured CBL intensities, the columns of $X$ are the basis functions defined by Eqs.~\eqref{eq:m1}, \eqref{eq:Mlin}, \eqref{eq:M10}, and
\eqref{eq:M13}, and $\boldsymbol{\beta}$ contains the corresponding coefficients. The coefficients are obtained by unconstrained least squares using the numerical pseudoinverse,
\begin{equation}
	\widehat{\boldsymbol{\beta}}=X^{+}\mathbf y .
	\label{eq:lstsq}
\end{equation}
The same measured points are used for all four models within each data set. Normalization used for visualization in Fig.~\ref{fig:allmaps} is applied only to the plotted maps; model fitting and model comparison use the measured intensities consistently within each data set.

Predictive performance is assessed with repeated random five-fold cross-validation. The points in each temperature/direction data set are partitioned into five folds. For each split, four folds are used to fit $\boldsymbol{\beta}$ and the remaining fold is predicted. Every fold is used once as the test set, and the complete five-fold procedure is repeated 20 times with new random partitions. For a held-out fold,
\begin{equation}
	R^2_{\mathrm{CV}}
	=1-
	\frac{\sum_i\left(I_i-\widehat I_i\right)^2}{\sum_i\left(I_i-\overline I_{\rm test}\right)^2}.
	\label{eq:cvr2}
\end{equation}
The reported $R^2_{\mathrm{CV}}$ is the mean of the fold-wise values over the 20 repeats; the repeat-to-repeat spread is retained for the supplementary numerical summary.

The four models contain different numbers of identifiable coefficients, so we also compare them using the Bayesian information criterion (BIC) \cite{Schwarz1978}. Using the least-squares residuals,
\begin{equation}
	\mathrm{BIC}=
	n\ln\!\left(\frac{\mathrm{RSS}}{n}\right)
	+k\ln n ,
	\label{eq:bic}
\end{equation}
where
\begin{equation}
	\mathrm{RSS}
	=\sum_{i=1}^{n}(I_i-\widehat I_i)^2
	\label{eq:rss}
\end{equation}
is evaluated from the fit to the complete data set, $n$ is the number of measured points, and $k$ is the numerical rank of the design matrix including the intercept. The effective parameter counts are $k=2$, 4, 4, and 9 for $M_1$, $M_{\rm lin}$, $M_{10}$, and $M_{13}$, respectively.
Complete numerical results are provided in Table~S1 of Supplement~1.

\FloatBarrier

\section{Results}
\label{sec-5}

\subsection{Test of Model $M_1$: circularity alone}

\begin{figure}[!htbp]
	\centering
	\includegraphics[width=\linewidth]{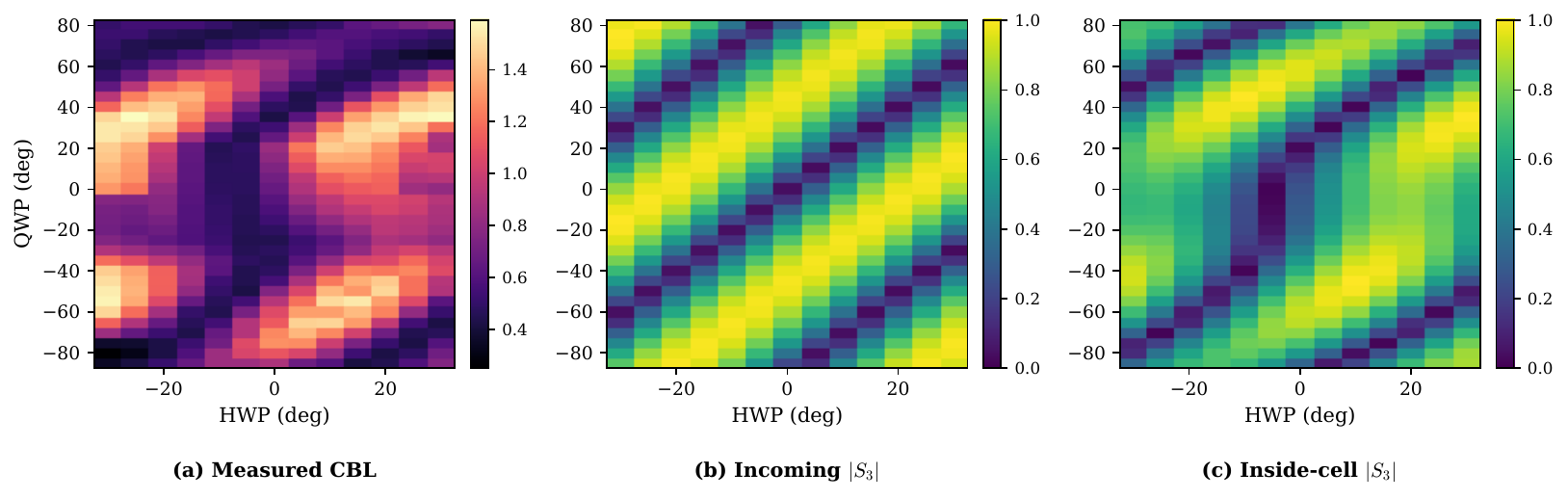}
	\caption{Representative circularity comparison for $T=100~^\circ$C and the $W_1\rightarrow W_2$ orientation. (a) Measured CBL, (b) $|S_3|$ for $\mathbf S_{\rm prep}$ before the cell, and (c) $|S_{3,\rm atom}|$ for the model-reconstructed 780-nm vapor-entrance state $\mathbf S_{\rm atom}$. The sapphire-window reconstruction substantially changes the circularity map, while the measured CBL contains additional structure captured by higher-dimensional Stokes descriptions.}
	\label{fig:challenge}
\end{figure}

We begin with the simplest question: can the CBL pattern be understood only from how circular the pump polarization is? If this were the case, the bright and dark regions of the CBL map should follow the map of $|S_3|$. Figure~\ref{fig:challenge} illustrates this test directly for a representative data set at 100~$^\circ$C in the $W_1\rightarrow W_2$ orientation.

The entrance sapphire window strongly changes where circular and nearly linear states occur, so the $|S_3|$ pattern calculated from $\mathbf S_{\rm atom}$ differs substantially from that of the externally prepared state $\mathbf S_{\rm prep}$. However, the measured CBL ridges and maxima still do not follow the reconstructed $|S_{3,\rm atom}|$ map.

The conclusion of this first test is therefore simple: correcting the polarization for the sapphire window is necessary, but circularity alone is not enough to describe the CBL response.

\FloatBarrier

\subsection{Measured maps and full quadratic response}

\begin{figure}[!htbp]
	\centering
	\includegraphics[width=\linewidth]{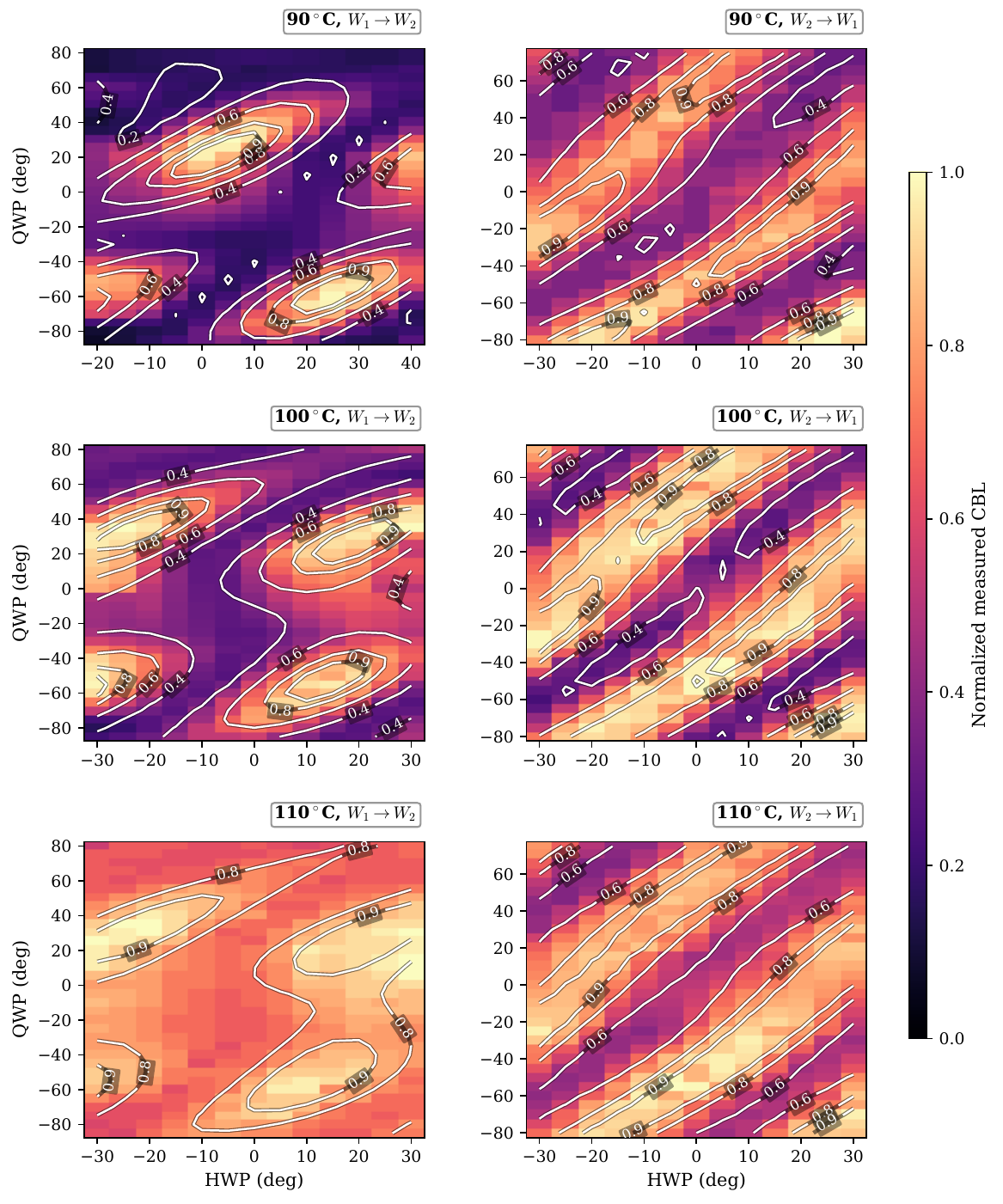}
	\caption{Measured total CBL-intensity maps with predictions of the complete second-order Stokes model $M_{13}$ evaluated from $\mathbf S_{\rm atom}$ for all three temperatures and both window orders. Each heat map is the measured CBL normalized to the maximum of that temperature/orientation set. White numbered curves are iso-level contours of the independently normalized $M_{13}$ prediction; the labels give the predicted level from 0.2 to 0.9.}
	\label{fig:allmaps}
\end{figure}

We next ask whether the measured CBL can be represented as a smooth function of the complete polarization state rather than of circularity alone. Figure~\ref{fig:allmaps} shows this comparison for all three temperatures and both window orders. The color map is the measured CBL, while the numbered white contours show the prediction of the full quadratic model $M_{13}$ calculated from $\mathbf S_{\rm atom}$. A contour marked 0.8, for example, corresponds to 80\% of the predicted maximum for that temperature and window order.

Although the detailed CBL pattern changes with temperature and when the cell is rotated, the predicted contours follow the main high- and low-intensity regions, including the position and curvature of the dominant ridges. In other words, once the actual polarization entering the vapor is reconstructed, the full Stokes state contains enough information to reproduce the main structure of the observed CBL maps with a compact empirical response surface.

The improvement is both structural and quantitative: the full-Stokes model reproduces the principal structure of the measured CBL response across the explored polarization space, including the locations and shapes of the dominant high- and low-signal regions while simultaneously providing substantially improved quantitative agreement.

\FloatBarrier

\subsection{Model comparison}

\begin{figure}[!htbp]
	\centering
	\includegraphics[width=\linewidth]{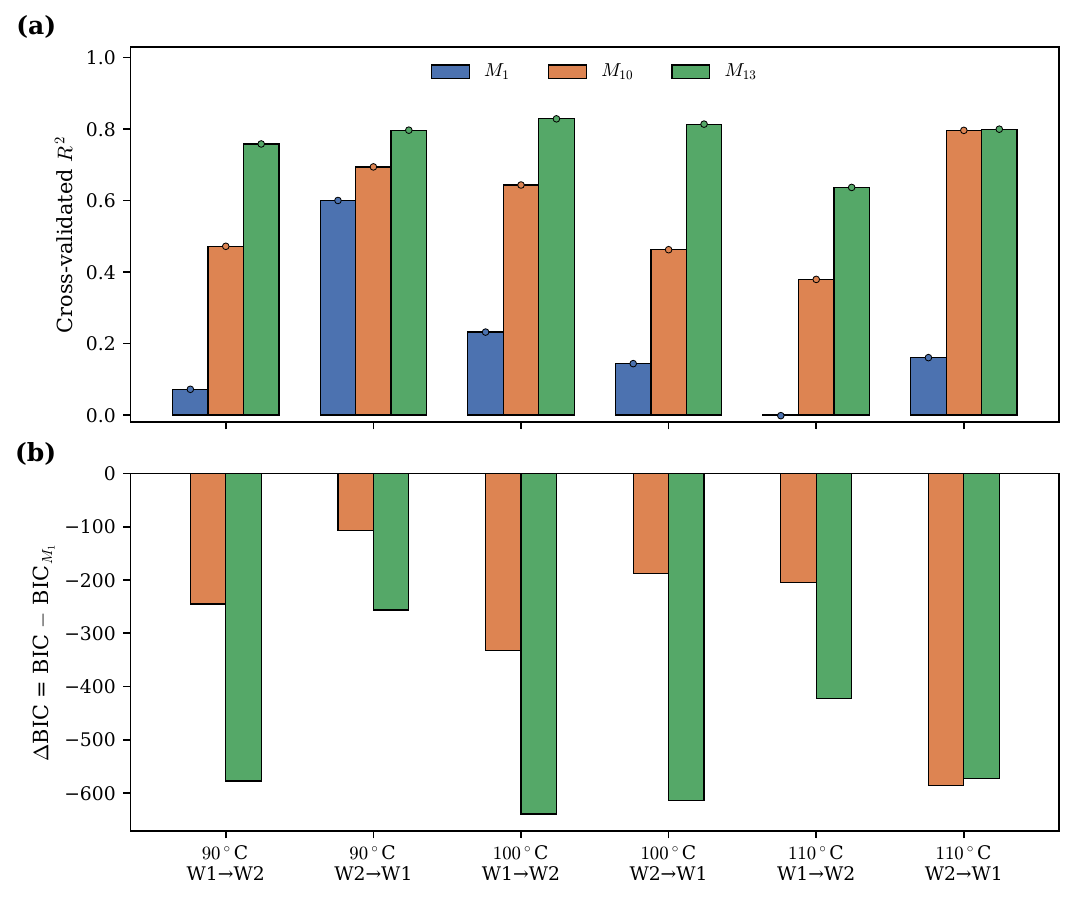}
	\caption{Comparison of the three empirical CBL response models. (a) Repeated five-fold cross-validated $R^2$ for the circularity-only model $M_1$, reduced model $M_{10}$, and full rank-9 quadratic model $M_{13}$. (b) $\Delta\mathrm{BIC}=\mathrm{BIC}-\mathrm{BIC}_{M_1}$ for $M_{10}$ and $M_{13}$; more negative values favor the indicated model after the parameter-count penalty is included.}
	\label{fig:metrics}
\end{figure}

A more flexible model will almost always fit the data better, so the visual agreement in Fig.~\ref{fig:allmaps} is not by itself sufficient.
Thus, we next compare the models using cross-validation, which tests their ability to predict CBL values at polarization settings excluded from the fit, and BIC, which evaluates the full-data fit while accounting for the different numbers of model coefficients.
Figure~\ref{fig:metrics} asks two practical questions: does each model predict points that were left out of the fit, and is the improvement large enough to justify the additional coefficients?

Panel (a) shows the repeated cross-validation result. 
Neither the circularity model $M_1$ nor the complete first-order model
$M_{\rm lin}$ provides a consistent description across the six data sets.
For $M_1$, the cross-validated $R^2$ ranges from $-0.002$ to 0.600,
whereas $M_{\rm lin}$ gives $-0.013$ to 0.147.
The reduced model $M_{10}$ improves this range to 0.379--0.796, showing that the linear-polarization content and selected mixed Stokes terms already matter. The full quadratic model $M_{13}$ gives fitted $R^2=0.656$--$0.839$ and cross-validated $R^2=0.636$--$0.828$ across the six data sets.

Panel (b) gives the complementary BIC comparison. Relative to $M_1$, $M_{13}$ gives $\Delta\mathrm{BIC}$ values between -639 and -256. It also has the lower BIC than $M_{10}$ in five of the six data sets. The exception is 110~$^\circ$C, $W_2\rightarrow W_1$, where $M_{10}$ is favored by 12.5 BIC units while the cross-validated scores of $M_{10}$ and $M_{13}$ are nearly identical (0.796 and 0.799, respectively). Thus, for this data set the additional quadratic terms add little predictive information.

Thus, these tests lead to the central model-comparison result of the paper: $M_1$ is too restrictive, $M_{10}$ captures a substantial part of the polarization dependence, and $M_{13}$ gives the most consistent overall description across temperatures and cell orientations. Complete numerical values are listed in Supplementary Table~S1.

\FloatBarrier

\section{Discussion}
\label{sec-6}

The sapphire windows transform the full polarization state entering the atomic vapor. We have developed a
calibrated optical model that provides a direct prediction of the polarization seen by the
atoms for any external wave-plate setting. 
The optical reconstruction and the CBL response fit are performered in two distinct stages. Off-resonant 780-nm polarimetry determines the effective complete-cell Jones transformation without using the CBL signal. With those optical parameters fixed, the factorization into entrance and exit operators supplies the model-reconstructed coordinate $\mathbf S_{\rm atom}$ used for the independent CBL data. The external analyzer measures the before-cell and complete-cell output Stokes states, while the internal vapor-entrance coordinate is generated by the fitted entrance-window operator.

The count of the parameters of the optical model reflects this global operator structure. Eight axis/rotation parameters are common to all three temperatures, and twelve retardances describe four effective operators at three temperatures. These 20 factor parameters act simultaneously on 266 full Stokes states, corresponding to 532 independent normalized Stokes coordinates. Table~S3 of Supplement~1 gives the fitted representative values, including the parameter values used for the trajectory in Fig.~\ref{fig:poincare}.
Because the external polarimetry determines the complete-cell transformation more directly than its decomposition into the two windows, we use one representative entrance/exit factorization to define a consistent reconstructed vapor-entrance coordinate, $\mathbf S_{\rm atom}$, for the CBL analysis.

The wavelength-resolved measurements provide a separate experimental test of the 780/776 chromatic dependence. They show closely matched states at the before-cell reference plane and a markedly larger wavelength transformation after the sapphire-window cell. 
It is shown, that only a small 780/776 mismatch at the before-cell reference plane exists,  while a substantially larger wavelength-dependent polarization transformation is observed after propagation through the assembled sapphire-window cell.
The six temperature--orientation data sets are treated independently, so each data set has its own measured best-fit rotation. Five data sets are represented with rms angular residuals below $8^\circ$, while the 110~$^\circ$C, $W_2\rightarrow W_1$ condition gives $12.87^\circ$. Table~\ref{tab:wavelength} therefore supplies both the chromatic rotation scale and the experimental accuracy of the rigid relation used in the two-pump reduction.

The analysis of CBL response uses the scalar total 420-nm detector signal as its dependent variable and a separate coefficient set for each temperature and orientation. The complete first-order model contains four coefficients and yields cross-validated $R^2$ values from $-0.013$ to 0.147. The complete second-order basis contains nine independent coefficients because the normalized-sphere identity removes one quadratic degree of freedom. Its cross-validated $R^2$ values of 0.636--0.828 show that the dominant smooth response structure in the present data is captured at second order. The reduced model $M_{10}$ further shows that selected quadratic coordinates already capture a large part of this structure.

A complete first-order or second-order scalar basis is closed under fixed
rotations of the normalized Stokes coordinate system. Consequently,
Eqs.~\eqref{eq:two-pump-linear-reduced} and
\eqref{eq:quadratic-reduced} show that a fixed relation between the
co-prepared pump trajectories preserves the complete first- and
second-order basis dimensions. The wavelength-resolved measurements
provide an experimental test of this rigid-trajectory representation
through the residuals reported in Table~\ref{tab:wavelength}. This
dimensionality argument is separate from the effective 780-nm
entrance/exit factorization and does not require a unique reconstruction
of the 776-nm vapor-entrance state.

The present Jones description represents normalized deterministic polarization transformations. The power-transmission ratios quoted in Sec.~\ref{sec-2} belong specifically to the off-resonant polarimetry condition and provide a propagation scale for that calibration.  More general input--output theories of dispersive and absorbing optical structures explicitly retain loss channels and non-unitary propagation \cite{Khanbekyan2003,Khanbekyan2005QED}. Resonant CBL generation involves its own pump attenuation, optical pumping, Doppler averaging, and phase-matching dynamics. A broader propagation model can incorporate those quantities together with spatially resolved diattenuation and beam steering when absolute conversion efficiency or microscopic susceptibility is the target observable.

\FloatBarrier

\section{Conclusions and outlook}
\label{sec-7}

We have measured and reconstructed the polarization dependence of coherent blue-light generation in a sapphire-window rubidium vapor cell containing natural Rb. 
The reconstruction is based on a full-Stokes model of the polarization transformation produced by the sapphire windows of the cell.
The sapphire windows produce a substantial and direction-dependent transformation of the externally prepared polarization, making full Stokes reconstruction necessary before the atomic response can be compared across experimental conditions.
We measured the external pump-polarization transformation of a sapphire-window rubidium cell at three temperatures and in two propagation orientations. A global 780-nm lossless Jones model reproduces 266 measured external output Stokes states with a $5.51^\circ$ rms angular mismatch. Direct before-cell polarimetry calibrates the preparation optics, direct after-cell polarimetry constrains the complete-cell transformation, and the effective factorization provides a model-reconstructed vapor-entrance coordinate.

Matched 780/776-nm Stokes measurements quantify the pump-wavelength dependence at every temperature and propagation orientation. The before-cell states are separated by $4.29^\circ$ on average, while the complete assembled cell gives temperature- and orientation-specific wavelength transformations corresponding to $22.20^\circ$--$32.02^\circ$ Poincar\'e rotations. The rigid-rotation rms residuals range from $4.83^\circ$ to $12.87^\circ$ and provide the experimental accuracy scale for representing the co-prepared two-wavelength trajectory by one fixed rotation for each temperature and orientation.

Using the calibrated 780-nm vapor-entrance coordinate, we analyze 2574 measurements of total spectrally filtered 420-nm CBL intensity. Across the six temperature--orientation data sets, the complete first-order Stokes model gives cross-validated $R^2=-0.013$--0.147, while the complete second-order model gives 0.636--0.828. 
In a two-pump quadratic coordinate model, a fixed specific rotation for given temperature and propagation direction
between the correlated pump trajectories reduces the response to the same
nine independent second-order coordinates. The directly measured
complete-cell 780/776 rotations and their residuals provide an experimental
test of the accuracy of this rigid-trajectory representation.
The measured wavelength residuals quantify the approximation involved in that reduction. This formulation keeps the external cell calibration, the reconstructed 780-nm intracell coordinate, and the empirical nonlinear-response fit as distinct steps.

Our model for predicting the polarization entering the vapor will be particularly useful for future CBL experiments with an applied magnetic field, where the optical polarization and Zeeman-state coupling become central control parameters \cite{Budker2002}.
The same reconstruction framework can be used for magnetic-field-dependent CBL measurements and a complementary experiment with dedicated 5.23~$\mu$m detection, polarization analysis, and correlation measurements with the simultaneously generated 420-nm field.

\clearpage
\backmatter

\bmsection{Funding}
This work was supported by the HESC of Armenia under Grant 22IRF-06.

\bmsection{Acknowledgment}
The authors thank A. Papoyan, A. Akulshin, D. Budker, R. Aramyan, and O. Tretiak for useful discussions and contributions to the development of this work.

\bmsection{Disclosures}
The authors declare no conflicts of interest.

\bmsection{Data Availability}
Data underlying the results presented in this paper are not publicly available at this time but may be obtained from the authors upon reasonable request.

\bmsection{Supplemental document}
See Supplement 1 for supporting content.

\bibliography{bibl}

\end{document}